\documentclass[11pt]{article}  

\usepackage{graphics,graphicx}
\usepackage{amsmath,amsfonts,amssymb}
\usepackage{algorithm}
\usepackage{algpseudocode}

\newcommand{\zb}{\mathbf{z}}

\newcommand{\lb}{\mathbf{\ell}}

\newcommand{\thB}{\mbox{\boldmath$\theta$}}
\newcommand{\thBs}{\mbox{\tiny \boldmath$\theta$}}

\newcommand{\ThB}{\mbox{\boldmath$\Theta$}}

\begin{document}


\title{Bayesian likelihood-free  localisation of a biochemical source using multiple dispersion models}

\author{Branko Ristic\thanks{Corresponding author: B. Ristic, DSTO, LD, Bld 94, Melbourne, Vic
3207, Australia; email: branko.ristic@dsto.defence.gov.au; tel: +61
3 9626 8370; fax: +61 3 9626 8473}, Ajith Gunatilaka, Ralph Gailis,
Alex Skvortsov\\Land Division\\Defence Science and Technology
Organisation\\ Australia }

\maketitle{}

\begin{abstract}
Localisation of a source of a toxic release of biochemical aerosols
in the atmosphere is a problem of great importance for public
safety. Two main practical difficulties are encountered in this
problem: the lack of knowledge of the likelihood function of
measurements collected by biochemical sensors, and the plethora of
candidate dispersion models, developed under various assumptions
(e.g. meteorological conditions, terrain). Aiming to overcome these
two difficulties, the paper proposes a likelihood-free approximate
Bayesian computation method, which simultaneously uses a set of
candidate dispersion models, to localise the source. This estimation
framework is implemented via the Monte Carlo method and tested using
two experimental datasets.
\end{abstract}



\section{Introduction}

The threat of accidental or deliberate release of toxic (bio or
chemical) aerosols into the atmosphere, has been well documented
\cite{kendal}.
Wind, as the dominant transport mechanism in
the atmosphere, 
can generate strong turbulent motion, causing the released aerosol
to disperse as a plume whose spread increases with the downwind
distance \cite{Arya1998}. For the sake of public safety, it is of
utmost importance to rapidly detect and localise a source of toxic
release so that the mitigation actions can be carried out promptly.
Ideally, the environment should be monitored continuously by a
network of spatially distributed sensors to measure the
concentration of toxic aerosols at various locations. This paper is
devoted to the problem of source localisation using the
concentration measurements collected by such a sensor network.

Two major difficulties are encountered in the described context of
source localisation. The first is the choice of the most suitable
dispersion model; the second is the lack of a precise and accurate
probabilistic description of concentration measurements. Both
difficulties are elaborated below.

A dispersion model describes, via mathematical equations, the
physical processes that govern the atmospheric dispersion of the
released material within the plume. The primary purpose of a
dispersion model is to calculate the mean concentration of emitted
material at given sensor locations. A plethora of dispersion models
are in use today \cite{holmes_morawska_06} to account for specific
weather conditions, terrain, source hight, etc. The problem of
selecting the most suitable dispersion model using statistical
signal processing techniques \cite{makalic_14} is largely neglected
by the atmospheric research community. The only somewhat relevant
reference is \cite{yee_model_sel}, which applies a single dispersion
(Lagrange stochastic) model under multiple source assumptions in
order to estimate the number of active sources of a toxic release.

Many references have been published on the topic of biochemical
source localisation, under an adopted suitable dispersion model. The
standard solutions are based on optimisation techniques, such as the
nonlinear least squares \cite{matthes_05},  that are known to fail
due to local minima or poor convergence. A version of the
least-squares is used in \cite{thomson_07} with minimisation carried
out via simulated annealing. The alternatives are Bayesian
techniques. Keats {et al.} \cite{keats_07} solved the chemical
source localisation problem for the case of a transient release in
the Bayesian framework using Markov chain Monte Carlo (MCMC) and
assuming Gaussian likelihood function of measurements. A similar
approach was adopted in \cite{humphries_12}. Bayesian framework with
Gaussian likelihood of measurements  was also applied in
\cite{ortner} to localise a biochemical source in urban
environments. A version of the MCMC was proposed in
\cite{senocak_08}, but assuming the log-normal likelihood function
of measurements.

The Bayesian approaches, such as
\cite{keats_07,humphries_12,ortner,senocak_08,yee_model_sel}, are
preferred than the optimisation techniques, because they result in
the posterior density function of the source location, thereby
providing an uncertainty measure to any point estimate derived from
it. However, the Bayesian approaches referenced above, require a
precise specification of the probabilistic model of the likelihood
function (e.g. Gaussian, log-normal, with their parameters).  Note
that errors in the measurements are not only due to sensor noise but
also due to the modeling inaccuracies, both of which in practice are
very difficult to specify precisely. A large part of the difficulty
arises from the fact that most approaches to source estimation have
involved the use of ensemble mean concentration models of
atmospheric tracer dispersion, but in reality, sensors are exposed
to widely fluctuating concentration  fields in both space and time,
that are also stochastically non-stationary.
This brings us to the second difficulty mentioned above: the lack of
precise and accurate probabilistic descriptions of concentration
measurements.

In this paper we develop a framework which overcomes both of the
aforementioned difficulties: a likelihood-free approximate Bayesian
computation (ABC) method for the localisation of a biochemical
source in the combination with the selection of the most suitable
model from a set of candidate dispersion models. This estimation
framework is implemented via the Monte Carlo method and tested using
two experimental datasets.

The paper is organised as follows. Section \ref{s:2} describes the
problem in a formal manner and also introduces a solution framework
in the form of the multiple-model ABC rejection sampler. Section
\ref{s:3} presents the adopted dispersion models for source
localisation. Section \ref{s:4} describes the proposed adaptive
iterative multi-model ABC sampler. Section \ref{s:num_ana} presents
the numerical results, obtained using the experimental datasets.
Finally, the main findings of this study are summarised in Section
\ref{s:6}.

\section{Background}
\label{s:2}

\subsection{Problem formulation}
\label{s:2.1}

Let us assume that the system (the atmospheric tracer dispersion)
obeys one of a finite number of models. A discrete random variable
$m$ denotes the model, whose domain $\mathcal{M}\subset \mathbb{N}$
has cardinality $M = |\mathcal{M}|\geq 1$. Each model
$m\in\mathcal{M}$ is parameterised by a vector $\thB_m \in\ThB_m$,
where $\ThB_m\subseteq\mathbb{R}^m$ is the corresponding
parameter space. The models, in general, are not nested
\cite{makalic_14}, although the intersection of  parameter spaces
$\ThB = \ThB_1\cap\dots\cap\ThB_M $ is non-empty; $\ThB$ represents
the space of the core parameters which includes the coordinates of
source location.

Let a concentration measurement from sensor $s=1,\dots,S$ (sensor
locations are known) be denoted $\zeta_s$. All sensor measurements
are stacked into a vector $\zb = [\zeta_1,\dots,\zeta_S]^\intercal$.

The problem is cast in the Bayesian framework. Let
$\pi_{\thBs_m}(\thB_m|m)$ denote the prior distribution over the
parameter space $\ThB_m$, with $m=1,\dots,M$. The posterior
distribution of the parameter vector $\thB_m$ follows from the Bayes
theorem:
\begin{equation}
p(\thB_m|m,\zb) =
\frac{\ell(\zb|\thB_m,m)\pi_{\thBs_m}(\thB_m|m)}{p(\zb|m)}
\label{e:post}
\end{equation}
where $\ell(\zb|\thB_m,m)$ is the likelihood function and
\begin{equation} p(\zb|m) = \int_{\ThB_m}
p(\zb|\thB_m,m)\pi_{\thBs_m}(\thB_m|m)d\thB_m
\end{equation}
is the marginal probability of measurement $\zb$ given model $m$.

Let $\pi_m(m)$ denote the prior distribution over the candidate
models $\mathcal{M}$. Then the posterior distribution over the
models is also obtained using the Bayes theorem as:
\begin{equation}
p(m|\zb) =
\frac{p(\zb|m)\pi_m(m)}{\sum_{m\in\mathcal{M}}p(\zb|m)\pi_m(m)}.
\end{equation}
Our ultimate goal is the posterior of the core parameter vector
$\thB\in\ThB$, which can be obtained via  model averaging
\cite{hoeting_99},
\begin{equation}
p(\thB|\zb) = \sum_{m\in\mathcal{M}}p(\thB_m|m,\zb)p(m|\zb),
\label{e:core1}
\end{equation}
where $p(\thB|m,\zb)$ is the posterior (\ref{e:post}) over the
subspace $\ThB\subseteq\ThB_m$.

\subsection{Approximate Bayesian computation}

ABC constitutes a class of Bayesian-type algorithms developed for
the estimation of a parameter vector in situations where the
likelihood function is intractable or unknown \cite{marin_12}. For
the case we consider, with multiple candidate models,  the
expressions for likelihoods $\ell(\zb|\thB_m,m)$, which feature in
(\ref{e:post}), are unknown.  ABC methods replace the unknown
likelihood $\ell(\zb|\thB_m,m)$ with the comparison between the
observed measurement and the measurement synthesized using the model
$m\in\mathcal{M}$. The simplest ABC algorithm is the ABC rejection
sampler \cite{pritchard_99}. In the context of parameter estimation
with multiple models, the ABC rejection sampler draws $N$ samples
from an approximation of the joint posterior
$p(\thB_m,m|\zb)=p(\thB_m|m,\zb)p(m|\zb)$. The pseudo-code of the
ABC rejection sampler for multiple models is presented in
Alg.\ref{a:1}.
\begin{algorithm}[tbhp]
 \caption{Multiple-model ABC rejection sampler} {
\begin{algorithmic}[1]
\State {\textbf{Input}}: $\zb$; $\epsilon$; $N$ \State Initialise:
$\mathcal{X}_1=\cdots=\mathcal{X}_M = \emptyset$ \Repeat \State Draw
$m^*$ from $\pi_m(m)$ \State Draw $\thB_m^*$ from
$\pi_{\thBs_{m^*}}(\thB_m)$ \State Simulate measurement $\zb^*$
using model $m^*$ and parameter  $\thB_m^*$ \State Compute distance
$d^* = D(\zb,\zb^*)$ \If {$d^* \leq \epsilon}$ \State
$\mathcal{X}_{m^*} = \mathcal{X}_{m^*} \cup \{\thB_m^*\}$ \EndIf
\Until {$\sum_{m=1}^M |\mathcal{X}_m| = N$}
 \State {\textbf{Output}}:  $\mathcal{X}_1,\cdots,\mathcal{X}_M$
\end{algorithmic}
} \label{a:1}
\end{algorithm}

Distance $d=D(\zb,\zb^*)$ between the measurement vector and the
synthesized data using model $m^*$ is compared to the tolerance
$\epsilon>0$ in line 8 of Alg.\ref{a:1}. The output of Alg.\ref{a:1}
is a posterior $p(\thB_m,m|d(\zb,\zb^*)\leq \epsilon)$ approximated
by $M$ sets of random samples:
\begin{equation}
\mathcal{X}_m = \{\thB_m^{(i)}\}_{1\leq i \leq
L_m},\hspace{.5cm}(m=1,\dots,M)  \label{e:five}
\end{equation}
such that $\sum_{m=1}^M L_m = N$. Using (\ref{e:five}) one can
approximate the joint posterior
$p(\thB_m,m|\zb)=p(\thB_m|m,\zb)p(m|\zb)$ with:
\begin{eqnarray}
p(m|\zb) & \approx & \frac{L_m}{N} \label{e:bb1}\\
p(\thB_m|m,\zb) & \approx & \frac{1}{L_m}\sum_{i=1}^{L_m}
\delta(\thB_m - \thB_m^{(i)}) \label{e:bb2}
\end{eqnarray}
where $\delta(x)$ is the Dirac delta function. The accuracy of
approximation improves with larger $N$ and smaller $\epsilon$.

\section{Dispersion models}
\label{s:3}

This section describes $M=3$  candidate dispersion models to be used
in source localisation. The first two models are based on the
Gaussian plume model \cite{kendal,Arya1998, macdonald_03} while the third
model is referred to as the stretch exponential model \cite{Huang1999, Kumar2010, skvortsov_yee_11}.

Gaussian plume models adopt a Gaussian distribution of the plume in
the vertical and horizontal directions under steady state
conditions. These models are a solution of the equation of tracer
transport with constant wind velocity (advection-diffusion
equation).
By convention, the wind velocity vector coincides with the $x$ axis,
while the spread of the plume in $y$ and $z$ directions is
determined by the respective standard deviations $\sigma_y$ and
$\sigma_z$, commonly referred to as the Pasquill-Gifford sigmas
\cite{Arya1998, macdonald_03}.

Consider a biochemical source of the release rate $Q_0$ located at
coordinates $(x_0,y_0,z_0)$.  According to the Gaussian plume model,
the mean concentration of the released material at the location of
the $s$th sensor $(x_s,y_s,z_s)$ is given by \cite{Arya1998, macdonald_03}
\begin{equation}
C_{s} = \frac{Q_0}{2\pi\sigma_{y_s}\sigma_{z_s} U}\;
e^{-\frac{(y_s-y_0)^2}{2\sigma_{y_s}^2}}\left[
e^{-\frac{(z_s-z_0)^2}{2\sigma_{z_s}^2}}+e^{-\frac{(z_s+z_0)^2}{2\sigma_{z_s}^2}}\right]\label{e:disper}
\end{equation}
if $x_s>x_0$ and zero otherwise. Notation $U$ stands for the mean
wind speed. Note that the Pasquill-Gifford sigmas, $\sigma_{y_s}$
and $\sigma_{z_s}$, in (\ref{e:disper}) are assigned the sensor
index $s$, because they are computed at coordinate $x_s$. The
simplest model for $\sigma_{y_s}$ and $\sigma_{z_s}$ is based on the
linear relationship with the downwind distance, i.e.
\begin{eqnarray}
\sigma_{y_s} & = & \sigma_0 + \frac{\sigma_v}{U}(x_s-x_0), \label{e:lin} \\
\sigma_{z_s} & = & \sigma_0 + \frac{\sigma_w}{U}(x_s-x_0).
\label{e:linz}
\end{eqnarray}
Explanation of the terms that feature in
(\ref{e:lin})-(\ref{e:linz}): $\sigma_0$ is the size of the source
(in units of length); $\sigma_v$ and $\sigma_w$ are environmental
parameters which account for the fluctuations in transverse and
vertical velocities, respectively (in units of velocity). The first
model  we adopt for source localisation (referred to as $m=1$) is
the Gaussian plume model (\ref{e:disper}) with the spreads given by
(\ref{e:lin}) and (\ref{e:linz}). The parameter vector for model
$m=1$ consist of $7$ parameters, that is:
\begin{equation}
\thB_1 = \left[\begin{matrix} x_0 & y_0 & z_0 & \sigma_0 & B &
\alpha & \beta\end{matrix}\right]^\intercal , \label{e:th1}
\end{equation}
where $B=Q_0/U$, $\alpha = \sigma_v/U$ and $\beta = \sigma_w/U$.

While the first dispersion model is very simple,  it is unable to
handle different canopy properties (average height, roughness,
porosity). Model $m=2$ is adopted to overcome this shortcoming. It
is also based on (\ref{e:disper}) and (\ref{e:linz}), but the spread
in $y$ direction, $\sigma_{y_s}$, is a nonlinear function of the
downwind distance from the source:
\begin{equation}
\sigma_{y_s} = \sigma_0 + \frac{\sigma_v}{U}\rho
\left(\frac{x_s-x_0}{\rho}\right)^\gamma.  \label{e:sig_y_nonl}
\end{equation}
The exponent $\gamma$ and the scale $\rho$ (usually referred to as
effective roughness) in (\ref{e:sig_y_nonl}) explicitly capture the
canopy characteristics. The exponent $\gamma$ may also change with
meteorological conditions; its theoretical value for turbulent
dispersion is $\gamma=1/2$. The parameter vector for model $m=2$
thus consists of $9$ parameters:
\begin{equation}
\thB_2 = \left[\begin{matrix} x_0 & y_0 & z_0 & \sigma_0 & B &
\alpha & \beta & \rho & \gamma\end{matrix}\right]^\intercal.
\label{e:th2}
\end{equation}

The Gaussian plume dispersion models have two well-known
deficiencies \cite{Arya1998}. The assumption of constant wind speed
does not even hold approximately for ground releases (i.e. $z_0=0$),
and so vertical wind speed profiles are usually employed, such as
the logarithmic wall-law from turbulent similarity theory, or the
commonly used power-law profile \cite{Csanady1973, Stockie2011}. Use
of these more physically realistic terms in the advection-diffusion
equation imply fundamentally different solutions that are not of a
Gaussian nature. More accurate analysis implies an ``effective''
plume convection velocity \cite{skvortsov_yee_11} that should be a
function of downstream distance $x_s$, but this increases
significantly the dimension of the parameter space. Also, under
different meteorological conditions the functional form given by
(\ref{e:disper}) can vary in the manner that expression for
$\sigma_{y_s}$ in (\ref{e:sig_y_nonl}) cannot fully capture.

The third dispersion model adopted for source localisation is the
stretch exponential (SE) model. This model is included because it is
capable, at least in theory, to overcome the aforementioned
limitations of the Gaussian plume model. The SE model is a solution
of the equation for tracer transport with a power-law wind velocity
and turbulent diffusivity profiles, and is therefore more general
that the Gaussian plume model. It enables an explicit, and rather
simple, parameterisation of various meteorological condition and
canopy characteristics.
The pure SE model is only strictly applicable for ground-level sources.
For $z > 0$, modified Bessel function solutions apply, which become
difficult to handle for inverse source modelling applications.


For the current paper, we take a leading asymptotic term from
these solutions, so that the mean concentration at the location of
the $s$th sensor is given by \cite{Huang1999, Kumar2010, skvortsov_yee_11}
\begin{equation}
C_{s} = \frac{B}{2\rho^2}\;\left(\frac{\rho}{x_s-x_0}\right)^{\tau}
\left[\exp\left\{-\frac{z_s^r-z_0^r}{\sigma_z^r}
\right\}+\exp\left\{-\frac{z_s^r+z_0^r}{\sigma_z^r} \right\}
\right]\; e^{-\frac{(y_s-y_0)^2}{2\sigma_{y_s}^2}} \label{e:disper1}
\end{equation}
where
\begin{eqnarray}
\sigma_{y_s} & = & \sigma_0 +  \alpha\rho
\left(\frac{x_s-x_0}{\rho}\right)^{1/2}\\
\sigma_{z_s} & = & \sigma_0 + \phi\rho
r^{2/r}\left(\frac{x_s-x_0}{\rho}\right)^{1/r}
\end{eqnarray}
with $r=1+2\mu$ and $\tau = 1/2+(1+\mu)/(1+2\mu)$ such that $0\leq
\mu\leq 1$. Parameters $B$,  $\alpha$ and $\rho$ have already been
defined. Parameter $\phi$ is similar to parameter $\beta$ in
(\ref{e:th1}) and (\ref{e:th2}). Parameter $r$ is introduced to
capture the variability of meteorological conditions. It defines the
functional form of the vertical concentration profile: for $r=1$,
this profile is exponential, while for $r=2$ it is Gaussian.
Parameter $\tau$ describes the mean concentration decay along the
plume centreline and captures the variability of meteorological
conditions (but is also affected by the type of canopy). According
to the SE model, $r$ and $\tau$ are related (both depend on $\mu$).
However, this relationship is valid only for an idealized flow with
a power-law profile over the flat underlying surface (in this case
$\mu$ is the exponent in the wind velocity profile). In order to
make the model more flexible, we adopt a relationship $\tau = \nu
+(1+\mu)/(1+2\mu)$, where $\nu$ is a free parameter whose prior
probability has the mean value equal to $1/2$. Equation
(\ref{e:disper1}) is the first term of an expansion of the exact
solution for the mean concentration at downwind distance, much
greater than the plume spread.

In summary, the SE model consists of $10$ parameters; its parameter
vector is specified as follows:
\begin{equation}
\thB_3 = \left[\begin{matrix} x_0 & y_0 & z_0 & \sigma_0 & B &
\alpha & \phi & \rho & \mu & \nu\end{matrix}\right]^\intercal.
\label{e:th3}
\end{equation}

As it was mentioned in Sec.\ref{s:2.1}, the parameter spaces of the
three models are not nested.

\section{Adaptive iterative multiple-model ABC sampler}
\label{s:4}

The ABC rejection sampler described in Alg.\ref{a:1} is very
inefficient due to its low acceptance rate. Several improvements of
the ABC rejection have been proposed in the single model case, such
as the ABC MCMC sampler \cite{marjoram_03} and a few versions of the
ABC sequential Monte Carlo (SMC) sampler \cite{sisson_07, toni_09,
delmoral_12}. Following \cite{toni_09}, we propose an iterative
Monte Carlo  multiple-model ABC sampler, whose basic steps are
described by Alg. \ref{a:2}. The key feature of this algorithm is
that it performs the ABC rejection scheme using a monotonically
decreasing sequence of tolerance levels $\epsilon_1>
\epsilon_2>\dots \epsilon_{\text{\tiny T}}\geq 0$, until the final
tolerance
 is reached. The sequence of tolerances $\epsilon_t$,
$t=1,2\dots,T$ is computed by the algorithm, from the measurement
vector $\zb$, hence the number of iterations $T$ is not know in
advance. For a given model $m$, sampling is initially carried out
from the prior $\pi_{\thBs_m}(\thB_m)$, followed by sampling from a
sequence of intermediate distributions $p(\thB_m|m,d(\zb,\zb^*)\leq
\epsilon_t)$,  $t=1,2,\dots$, which gradually approach the target
distribution $p(\thB_m|m,d(\zb,\zb^*)\leq \epsilon_{\text{\tiny
T}})$. The theoretical justification of the proposed iterative
scheme, presented in \cite{toni_09}, is based on the sequential
importance sampling (SIS) paradigm.

There are two differences between our proposed Alg. \ref{a:2} and
the algorithm reported in \cite{toni_09}, both of which reflect the
adaptive nature of the former: (i) our algorithm does not require
the sequence of tolerances
$\epsilon_1,\epsilon_2,\cdots,\epsilon_{\text{\tiny T}}$ to be
specified as an input (it works it out from the data $\zb$); (ii)
The proposal distribution for each sample and at each iteration is
computed adaptively, in a manner similar to population Monte Carlo
techniques \cite{cappe_pMC}.

\begin{algorithm}[tbhp]
 \caption{:\hspace{3mm} Adaptive iterative multiple-model ABC rejection sampler} {
\begin{algorithmic}[1]
\State {\textbf{Input}}: $\zb$; $\Delta$; $N$; \State
$[\{\mathcal{X}^0_m\}_{1\leq m\leq M},\epsilon_1]$ =
Init-Iter$(\zb,N)$ \State $t=0$; $\epsilon_0 \leftarrow \infty$
\While {$(\epsilon_{t}- \epsilon_{t+1}) > \Delta$} \State $t = t+1$
\State $[\{\mathcal{X}^{t}_m\}_{1\leq m \leq M},\epsilon_{t+1}]$ =
Repeated-Iter$(\{\mathcal{X}^{t-1}_m\}_{1\leq m \leq
M},\epsilon_t,\zb,N)$ \EndWhile
 \State {\textbf{Output}}:  $\{\mathcal{X}^{t}_m\}_{1\leq m \leq M}$
\end{algorithmic}}
\label{a:2}
\end{algorithm}

 The initial iteration of  Alg.
\ref{a:2} (line 2) creates $M$ initial sample sets
$\{\mathcal{X}^0_m\}_{1\leq m\leq M}$ from the priors $\pi_m$,
$\pi_{\thBs_m}$, $m=1,\dots,M$, and computes the first tolerance
level $\epsilon_1$. The repeated iteration, line 6, performs the
rejection sampling at a given tolerance and also computes the
tolerance level for the next iteration. The ``while loop'' (lines
4-7) is terminated when the difference between the two consecutive
tolerances is below a certain threshold $\Delta$.

Since the algorithm is based on SIS, both the initial and the
subsequent samples $\{\mathcal{X}^t_m\}_{1\leq m\leq M}$, $t=0,1,..$
are {\em weighted}. Hence (\ref{e:five}) now takes the form:
\begin{equation}
\mathcal{X}^t_m =
\left\{\left(w_m^{(i,t)},\thB_m^{(i,t)}\right)\right\}_{1\leq i \leq
L^t_m},\hspace{.5cm}(m=1,\dots,M; t=0,1,\dots)  \label{e:nn}
\end{equation}
where importance weights $w_m^{(i,t)}$ are normalised, that is
$\sum_{i=1}^{L_m^t} w_m^{(i,t)} = 1$.

The steps of the initial iteration are given by Alg.\ref{a:3}. The
tolerance $\epsilon_1$ is computed in line 10 as an order statistic
$\varphi$ of sample distances $d_1,\cdots,d_N$.

\begin{algorithm}[tbhp]
 \caption{:\hspace{3mm} $[\{\mathcal{X}^0_m\}_{1\leq m\leq M},\epsilon_1]$ =
 Init-Iter$(\zb,N)$}
  {
\begin{algorithmic}[1]
\State {\textbf{Input}}: $\zb$;  $N$ \State $\mathcal{X}^0_1 =
\cdots = \mathcal{X}^0_M = \emptyset$ \For {$i=1,\dots,N$} \State
Draw $ m \sim \pi_{m}$ \State Draw $\thB^* \sim \pi_{\thBs_{m}}$
\State Simulate measurement $\zb^*$ using model $m$ with parameter
$\thB^*$ \State Compute distance $d_i = D(\zb,\zb^*)$ \State
$\mathcal{X}^0_{m} = \mathcal{X}^0_{m} \cup \{(w=1,\thB^*)\}$
\EndFor \State $\epsilon_1 = \varphi(d_1,\dots,d_N)$ \For
{$j=1,\dots,M$} \State Normalise weights in $\mathcal{X}^0_j$
\EndFor
 \State {\textbf{Output}}:  $\mathcal{X}^0_1, \cdots,
\mathcal{X}^0_M $, $\epsilon_1$
\end{algorithmic}}
\label{a:3}
\end{algorithm}

Finally, the pseudo-code of the repeated iteration  is given in
Alg.\ref{a:4}. The while-loop between lines 4 to 17 carries out
rejection sampling. The model $m$ is drawn from the prior in line 5,
followed by the selection of the sample index $k$ in line 7. This
selection is based on the weights in $\mathcal{X}^{t-1}_m$.  A
candidate sample $\thB^*$ is drawn from the proposal distribution
$q_m(\theta_m|\thB^{(k,t-1)}_m)$ in line 8. The proposal is adopted
as a normal distribution, whose mean is $\thB^{(k,t-1)}_m$ and the
covariance matrix is related to the sample covariance of the sample
$\mathcal{X}^{t-1}_m$. The practical implementation of line 8 is as
follows: $\thB^* = \thB^{(k,t-1)}_m + \lambda h \Sigma \varepsilon$,
where $0<\lambda<1$ is a parameter of the algorithm, $h$ is the
optimal bandwidth of the multivariate Gaussian kernel
\cite[Eq.(12.2.7)]{musso_et_al_00}, $\Sigma$ is the square-root of
the empirical covariance matrix of $\mathcal{X}^{t-1}_m$ and
$\varepsilon$ is a sample from the standard normal distribution. The
computation of an unnormalised weight for an accepted sample in line
12, is carried out in the manner of \cite{toni_09}. The output
tolerance $\epsilon_{t+1}$ is again computed as an order statistic
$\varphi$ of the sample $d_1,\dots,d_N$.

\begin{algorithm}[tbhp]
 \caption{:$[\{\mathcal{X}^{t}_m\}_{1\leq m \leq M},\epsilon_{t+1}]$ =
Repeated-Iter$(\{\mathcal{X}^{t-1}_m\}_{1\leq m \leq
M},\epsilon_t,\zb,N)$} {
\begin{algorithmic}[1]
\State {\textbf{Input}}: $\{\mathcal{X}^{t-1}_m\}_{1\leq m \leq M}$,
$\epsilon_t$, $\zb$, $N$ \State $\mathcal{X}^t_1 = \cdots =
\mathcal{X}^t_M = \emptyset$
 \State
$n=0$ \While {$n<N$}  \State Draw $m\sim \pi_{m}$ \State $L_m =
|\mathcal{X}^{t-1}_m|$ \State Select index $k\in\{1,\dots,L_m\}$
with $\mathbb{P}(k=j)=w^{(j,t-1)}_m$ \State Draw $\thB^* \sim
q_m(\cdot|\thB^{(k,t-1)}_m)$ \State Simulate measurement $\zb^*$
using model $m$ with $\thB^*$ \State Compute distance $d^* =
D(\zb,\zb^*)$ \If  {$d^* \leq \epsilon_t$} \State Weight
$\tilde{w}^* = \frac{\pi_{\thBs_m}(\thB^*)}{\sum_{i=1}^{L_m}
w_m^{i,t-1}\,q_{m}(\thB^*|\thB^{(i,t-1)}_m)}$ \State Accept:
$\mathcal{X}_m = \mathcal{X}_m \cup \{(\tilde{w}^*,\thB^*)\}$ \State
$n=n+1$ \State $d_n = d^*$ \EndIf \EndWhile \State $\epsilon_{t+1} =
\varphi(d_1,\cdots,d_N)$ \For {$m=1,\dots,M$} \State Normalise
weights in $\mathcal{X}^{t}_m$ \EndFor
 \State {\textbf{Output}}:  $\{\mathcal{X}^{t}_m\}_{1\leq m \leq M}$, $\epsilon_{t+1}$
\end{algorithmic}}
\label{a:4}
\end{algorithm}

Recall that out goal is to localise the source. Denote by vector
$\lb = [x_0\;\;y_0]^\intercal \in\mathbb{L}$ the coordinates of the
source. The space of source coordinates is clearly a subspace of the
core parameter space. From the output of Alg.\ref{a:2}, expressed by
(\ref{e:nn}) at the last iteration $t=T$, we can extract  $M$ sets
of random samples over the subspace $\mathbb{L}$:
\begin{equation}
\mathcal{L}^t_m =
\left\{\left(w_m^{(i,t)},\lb_m^{(i,t)}\right)\right\}_{1\leq i \leq
L^t_m},\hspace{.5cm}(m=1,\dots,M; t=T).  \label{e:nnl}
\end{equation}
According to (\ref{e:core1}), the posterior density of source
location is then approximated by $\mathcal{L}^t_m$ as follows:
\begin{equation}
p(\lb|\zb) \approx \sum_{m=1}^M \frac{L_m}{N} \sum_{i=1}^{L_m}
w_m^{(i,t)}\,\delta(\lb - \lb_m^{(i,t)})  \label{e:post_l}
\end{equation}

\section{Numerical analysis}
\label{s:num_ana}

\subsection{Experimental datasets}

Algorithm evaluation was carried out using two experimental datasets
collected by COANDA Research \& Development Corporation. The
experiments were carried out  using their large recirculating water
channel, specially designed for dispersion modelling. The water
channel is 10 m long, 1.5 m wide and 0.9 m deep. The floor of the
water channel was covered with a metal mesh of height 4 mm to give
surface roughness.

The source was releasing fluorescein dye, at a constant rate,  from
a narrow vertical tube, placed $z_0=4$ mm above the bottom of the
channel at coordinates $x_0=-373.5$ mm, $y_0=0$ mm. Concentration
data were collected at several downstream positions using 1D laser
induced fluorescence linescan system, at the rate of 300 lines per
second, for a total sampling time of 1000 seconds~\cite{yee06}.

\begin{figure}[htbp]
\centerline{\includegraphics[height=5.2cm]{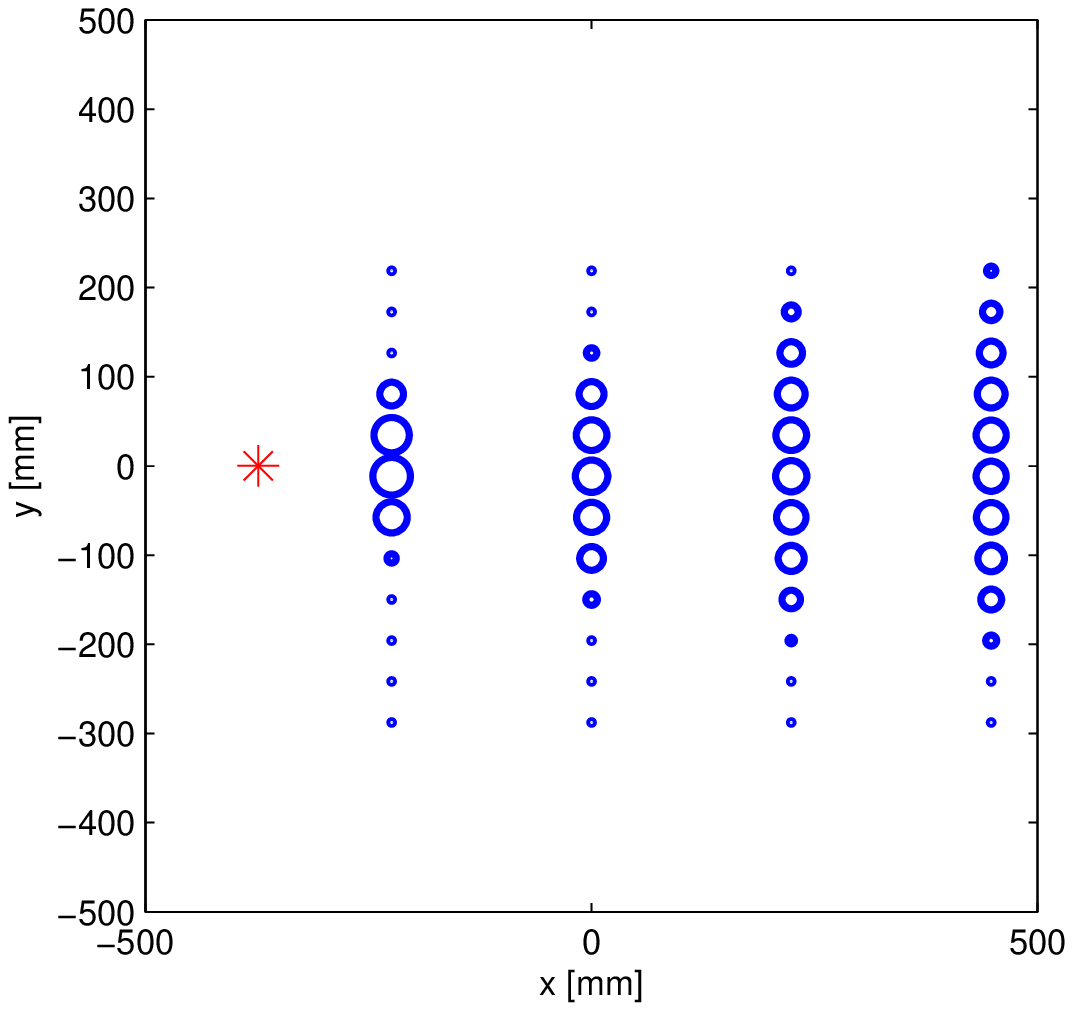}
\includegraphics[height=5.2cm]{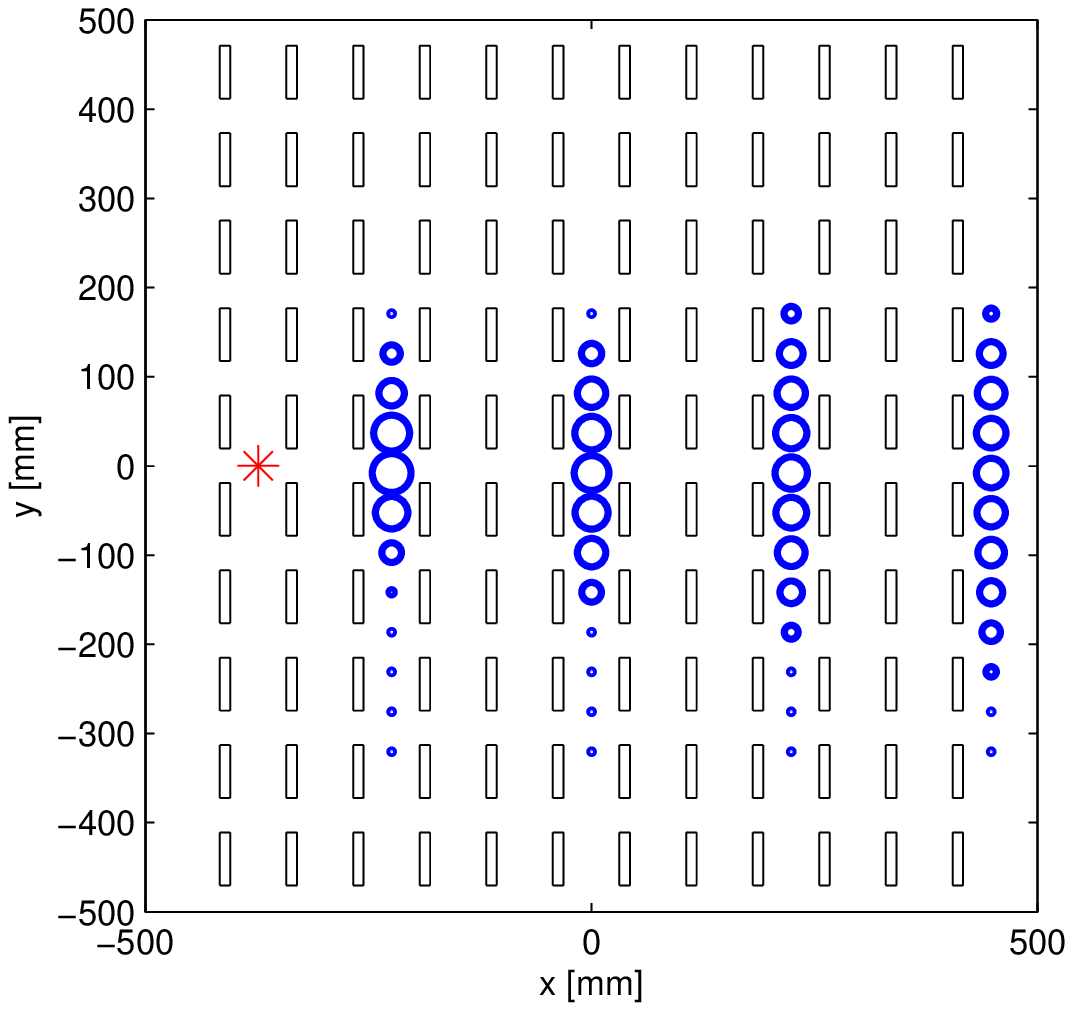}}
\centerline{(a)\hspace{6.5cm}(b)}
 \caption{Top-down view of the experimental setup: (a) dataset 1,  (b) dataset 2.  Location of the source
 is marked by a red asterisk at
 $(-373.5,0)$. Sensor locations indicated by blue
 circles, whose radius is proportional (on the log-scale) to the corresponding concentration measurement. The rectangles in (b)
 indicate the contours of the obstacles}
 \label{f:setup1}
\end{figure}

Two experimental datasets are used for algorithm evaluation (both
are available as supplementary material of this submission). Dataset
1 was collected in the absence of any obstacles (mimicking an open
terrain scenario). Dataset 2 was collected in the presence of 10 mm
high obstacles, placed on a regular grid, thus mimicking an urban
scenario. Both datasets were  extracted from the full recordings
averaged over 100 seconds, and consist of $S=48$ sensor measurements
(four rows of 12 sensors) at downstream positions. The top-down view
of the experimental setup for both cases is shown in
Fig.\ref{f:setup1}. The source location at  coordinates $(-373.5,0)$
is marked by a red asterisk. The position of sensors is indicated by
blue circles whose radius is proportional (on the log-scale) to the
corresponding concentration measurement. The height of all sensors
in both setups was $z_s=9.3$ mm, for $s=1,\dots,48$.

\subsection{Priors and parameters}

We adopted the prior over $M=3$ models to be uniform, that is
$\pi_m(1)=\pi_m(2) = \pi_m(3) = 1/3$.

The priors for various parameters included in vectors $\thB_1$,
$\thB_2$ and $\thB_3$, specified by (\ref{e:th1}), (\ref{e:th2}) and
(\ref{e:th3}), respectively, were adopted as follows (all units of
length are {\em millimeters}):
\begin{equation*}
\begin{aligned}
 \pi(x_0)  & = \mathcal{U}(-1000, 0), \\
 \pi(z_0)  & =  \mathcal{G}(1.333, 3), \\
\pi(B) & =  \mathcal{G}(2, 2.5), \\
\pi(\beta)  & =  \mathcal{G}(1.667,0.15),\\
\pi(\phi)  & =  \mathcal{G}(1.667,0.15),\\
\pi(\mu)  & =  \mathcal{B}(1.5, 3),
\end{aligned}\hspace{.5cm}
\begin{aligned}\pi(y_0)  & =   \mathcal{U}(-500, 500),\\
 \pi(\sigma_0) & =  \mathcal{G}(15.5,0.03),\\
 \pi(\alpha)  & =  \mathcal{G}(3, 0.5), \\
\pi(\rho) & = \mathcal{G}(6,1), \\
\pi(\gamma)  & =  \mathcal{B}(3,3),\\
 \pi(\nu) & = \mathcal{B}(6,6).
\end{aligned}
\end{equation*}
Here  $\mathcal{U}(a, b)$ is the uniform distribution, with limits
$a$ and $b$,  $\mathcal{G}(k, \eta)$ is the Gamma distribution with
shape $k$ and scale $\eta$, and $\mathcal{B}(p,q)$ is the Beta
distribution with parameters $p$ and $q$.

The proposed  ABC sampler described by Alg.\ref{a:2}, was executed
using $N=1000$ samples. The distance $D$ between the actual
measurement $\zb = [\zeta_1,\dots,\zeta_S]^\intercal$ and the
synthesised ``measurement'', using model $m$, denoted $\zb^* =
[C_1(\thB_m),\dots,C_S(\thB_m)]^\intercal$, was adopted as:
\begin{equation}
d(\zb,\zb^*)=\sum_{s=1}^S (\zeta_s - C_s(\thB_m))^2.
\end{equation}
Statistics $\varphi$, which computes the tolerance level for the
next iteration, was adopted to be the $128$th smallest value of
samples $d_1,\dots,d_{N=1000}$. The parameter $\lambda$, used in the
proposal $q_m$, was set to $\lambda=0.4$. The termination threshold
was adopted as $\Delta = 2\cdot 10^{-10}$.

\subsection{Results}

\begin{figure}[htbp]
\centerline{\includegraphics[height=5.cm]{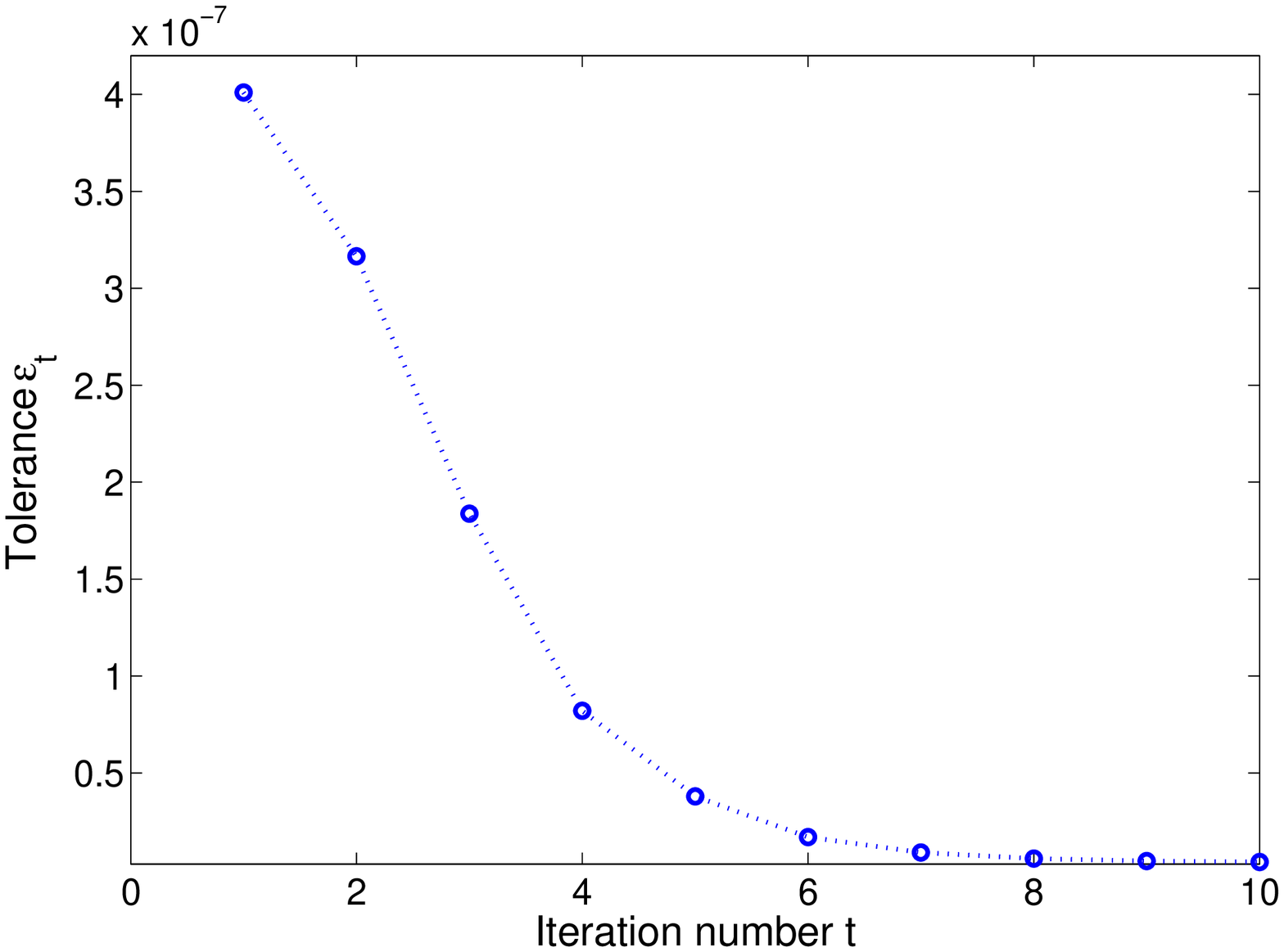}
\includegraphics[height=5.cm]{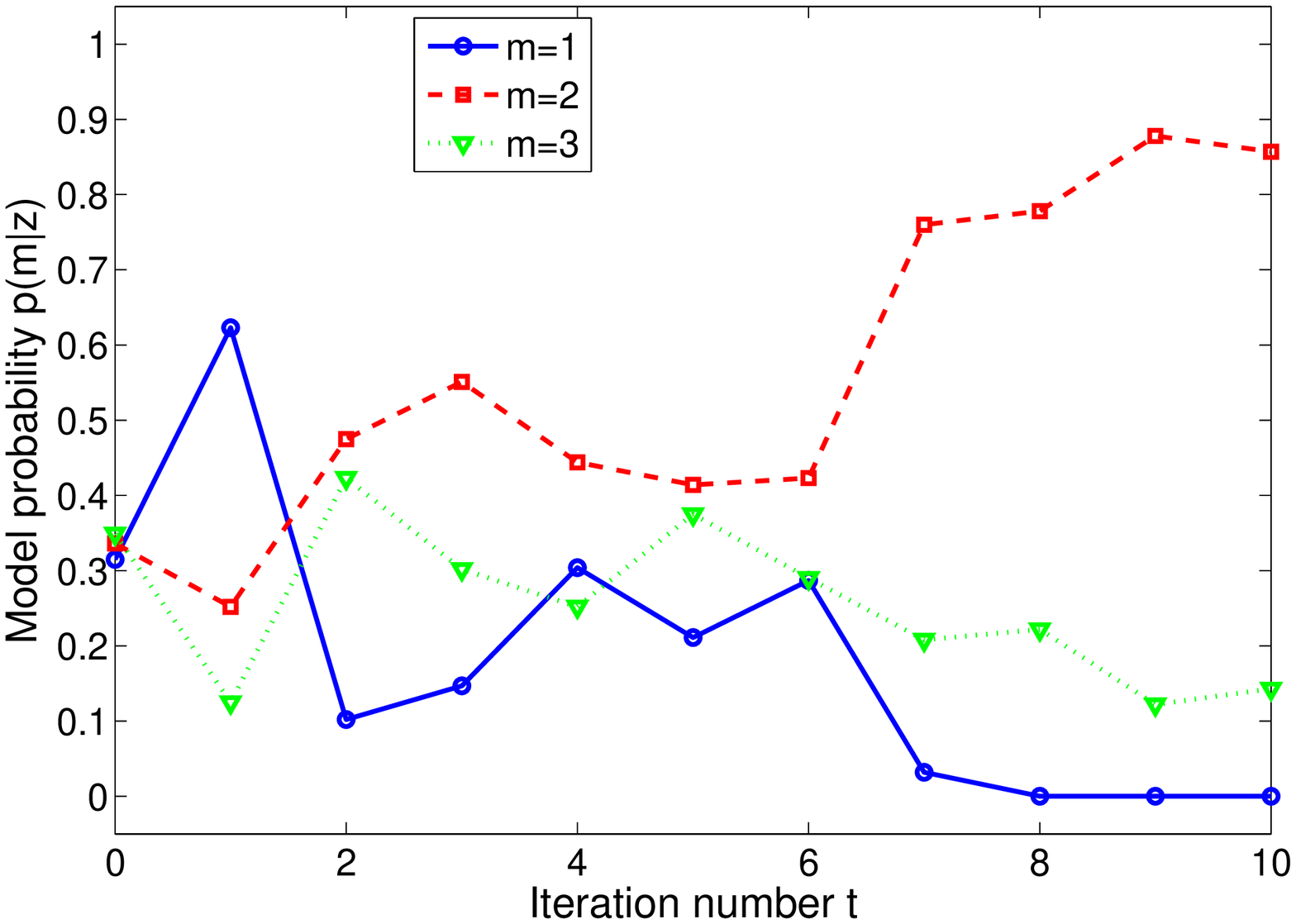}} \centerline{(a)\hspace{6cm}(b)}
 \caption{Tolerance levels and model probabilities over iterations (dataset 1): (a) $\epsilon_t$; (b)
 $p(m|\zb)$. }
 \label{f:2}
\end{figure}

The results for dataset 1 are presented first.  The tolerance levels
computed by the proposed ABC sampler are shown in Fig.\ref{f:2}.(a).
It took $t=10$ iterations to reach the final tolerance level of
$\epsilon_{\text{\tiny T}} = 4.04\cdot 10^{-9}$.  The average
acceptance rate, over all iterations, was $2.8\%$. The model
probabilities, $p(m|\zb)$ are shown at each iteration in
Fig.\ref{f:2}.(b). After the initial iteration  ($t=0$), the
probabilities of all three  models are approximately $1/3$.
Subsequently, while the tolerance levels were high, the three
probabilities fluctuate until the iteration $t=7$. After that, the
probability of model $m=1$ drops to zero, while the probability of
models $m=2$ grows to about 0.85. The probability of model $m=3$
drops, but never goes to zero.

\begin{figure}[htb]
\centerline{\includegraphics[height=8cm]{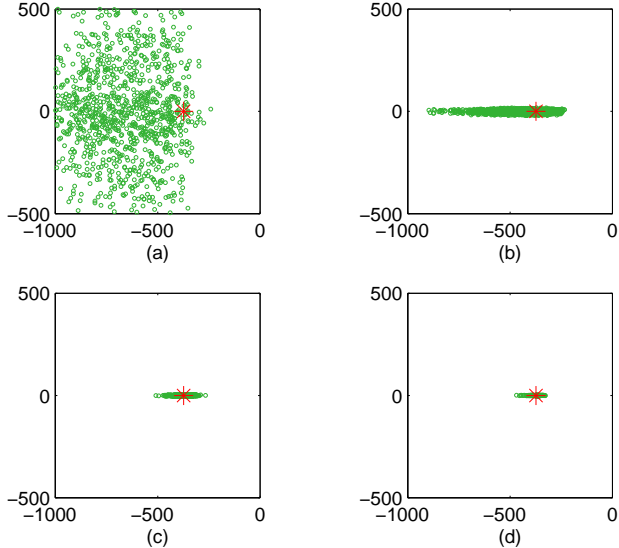}}
 \caption{Scatter plots of random samples $\mathcal{L}^t_m$ in the $(x,y)$ plane (dataset 1). The samples approximate the posterior density
$p(\lb|\zb)$. Scatter plots shown after iteration: (a) $t=1$, (b)
$t=4$, (c) $t=7$, (d) $t=10$. The true source location is marked
with a red asterisk. }
 \label{f:3}
\end{figure}

Fig.\ref{f:3} shows the scatter plot of random samples
$\mathcal{L}^t_m$ which approximate the posterior density
$p(\lb|\zb)$ of (\ref{e:post_l}),  at iterations (a) $t=1$, (b)
$t=4$, (c) $t=7$ and (d) $t=10$. The true source location is marked
by a red asterisk. Localisation of the source clearly improves with
iterations, as the tolerance levels get smaller.

\begin{figure}[htbp]
\centerline{\includegraphics[height=4.5cm]{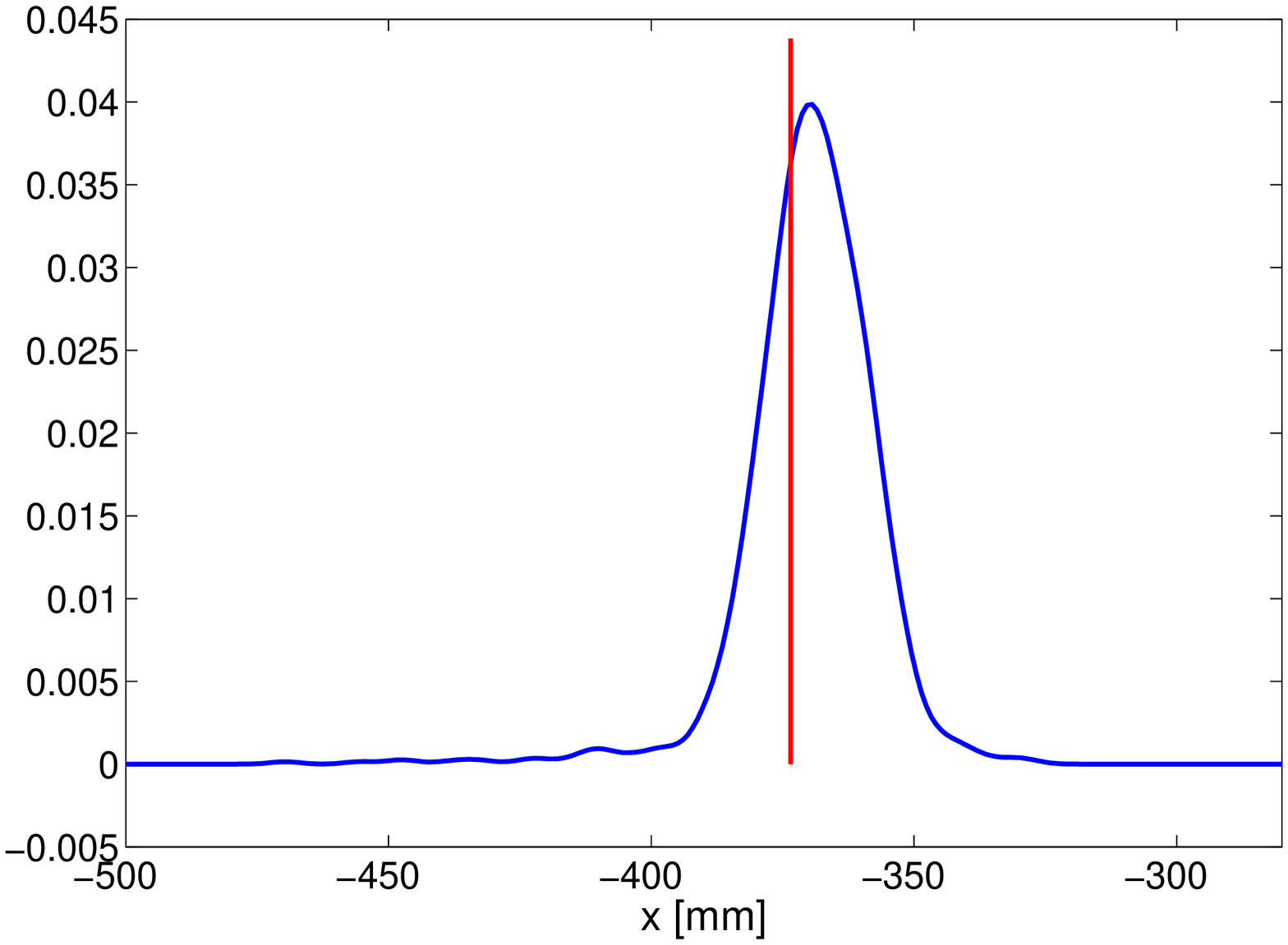}
\includegraphics[height=4.5cm]{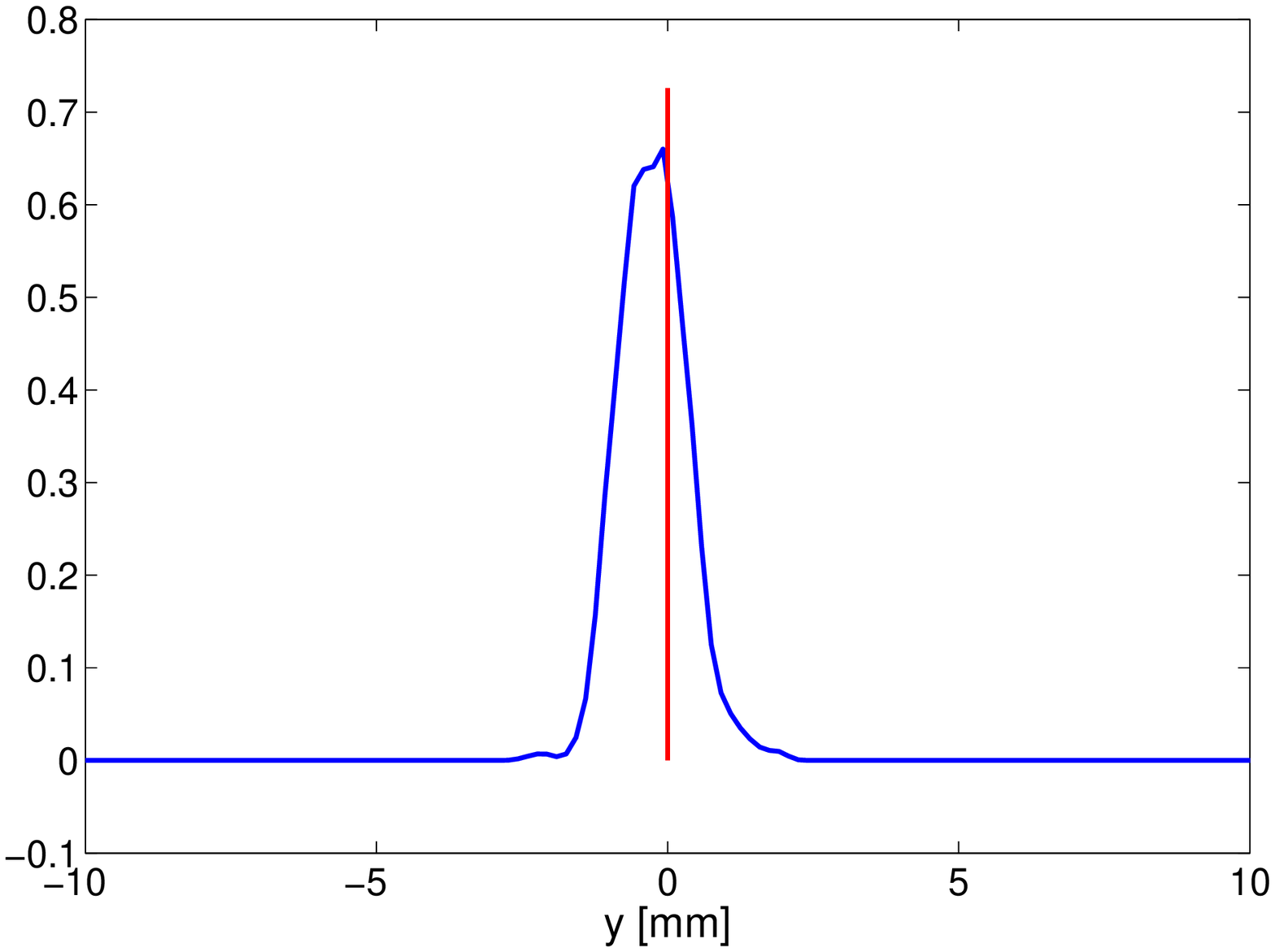}} \centerline{(a)\hspace{6cm}(b)}
 \caption{Estimated posterior density of the source location (dataset 1): (a) $x$ axis, (b) $y$ axis.
 The true values are indicated by the vertical solid lines. }
 \label{f:4}
\end{figure}

Finally, Figs.\ref{f:4}.(a) and (b) display the final posterior
distribution after the iteration $t=10$, marginalised to $x$, and
$y$ axes, respectively. The density estimates from the random sample
are obtained using the kernel density estimation (KDE)
\cite{silverman_86}. The true source coordinates are marked by the
solid vertical lines.

The first observation is that the estimated marginal posteriors are
remarkably accurate (considering that the prior was a uniform
density with the span of 1000 mm) and include the true source
location. The second observation is that the estimation is
significantly better for the coordinate $y_0$, than for $x_0$, which
is in accordance with the theoretical analysis carried out using the
Carmer-Rao bound \cite{ssp14_crb}.

\begin{figure}[htbp]
\centerline{\includegraphics[height=5.cm]{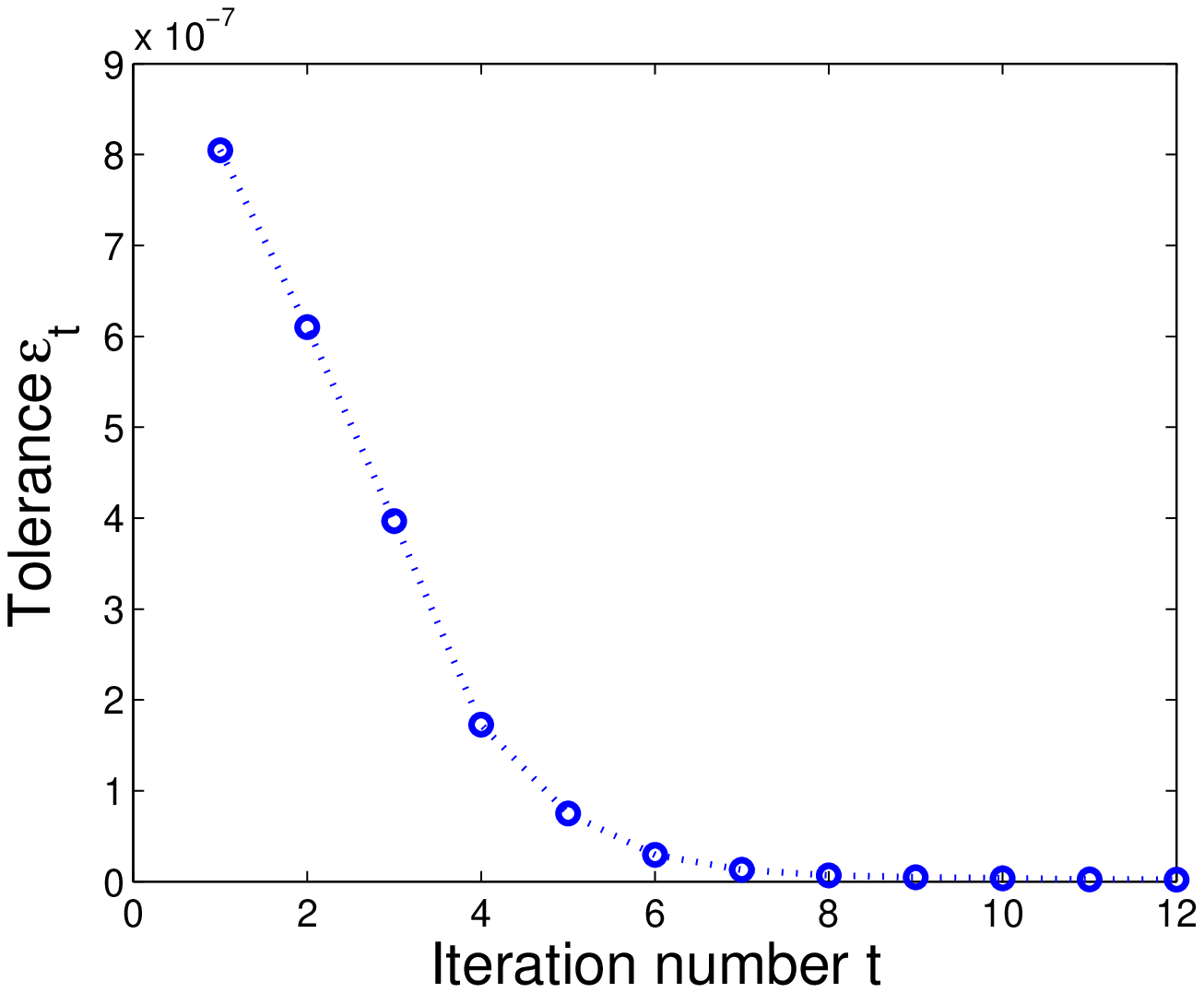}
\includegraphics[height=5.cm]{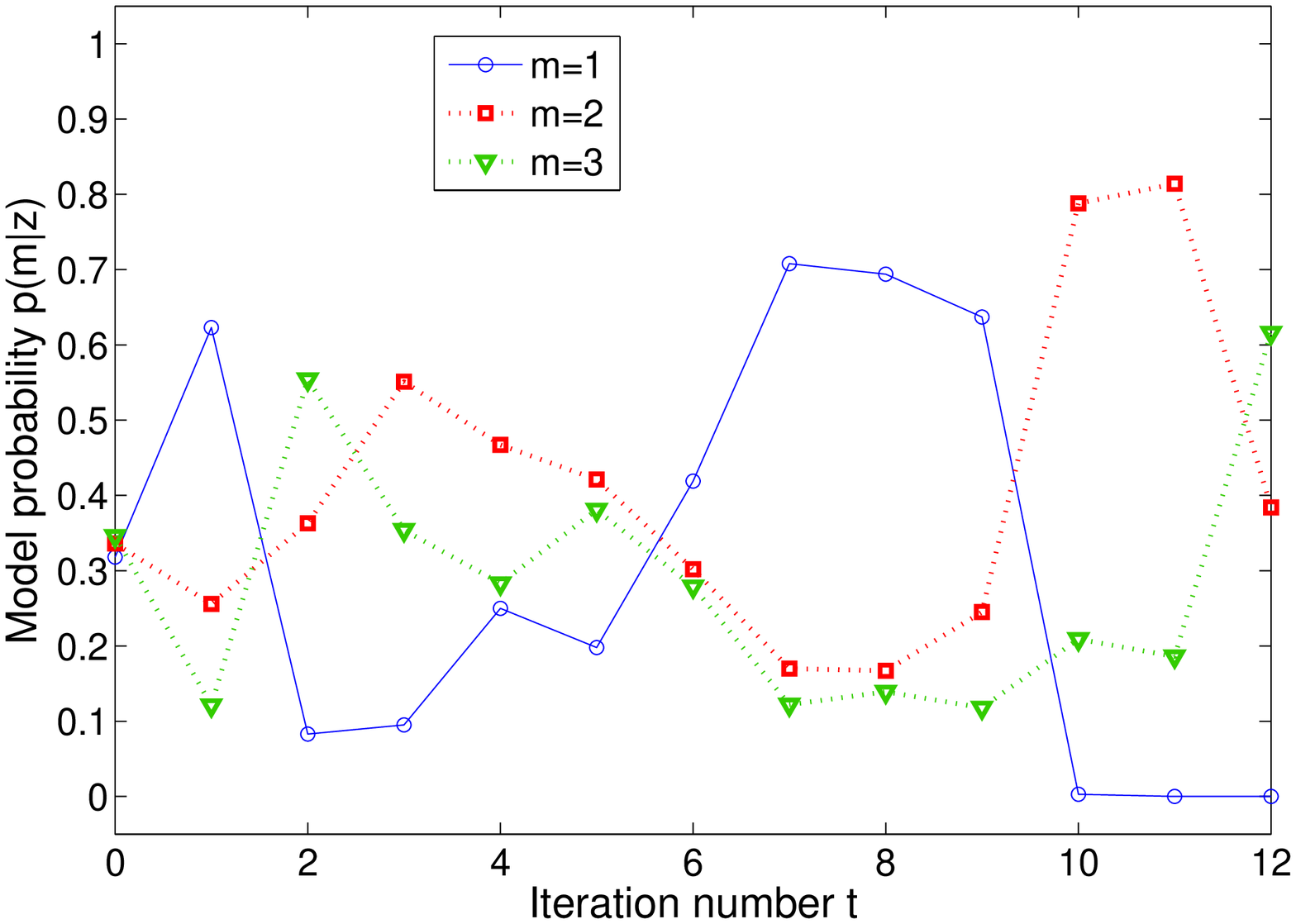}} \centerline{(a)\hspace{6cm}(b)}
 \caption{Tolerance levels and model probabilities over iterations (dataset 2): (a) $\epsilon_t$; (b)
 $p(m|\zb)$. }
 \label{f:5}
\end{figure}

\begin{figure}[htb]
\centerline{\includegraphics[height=8cm]{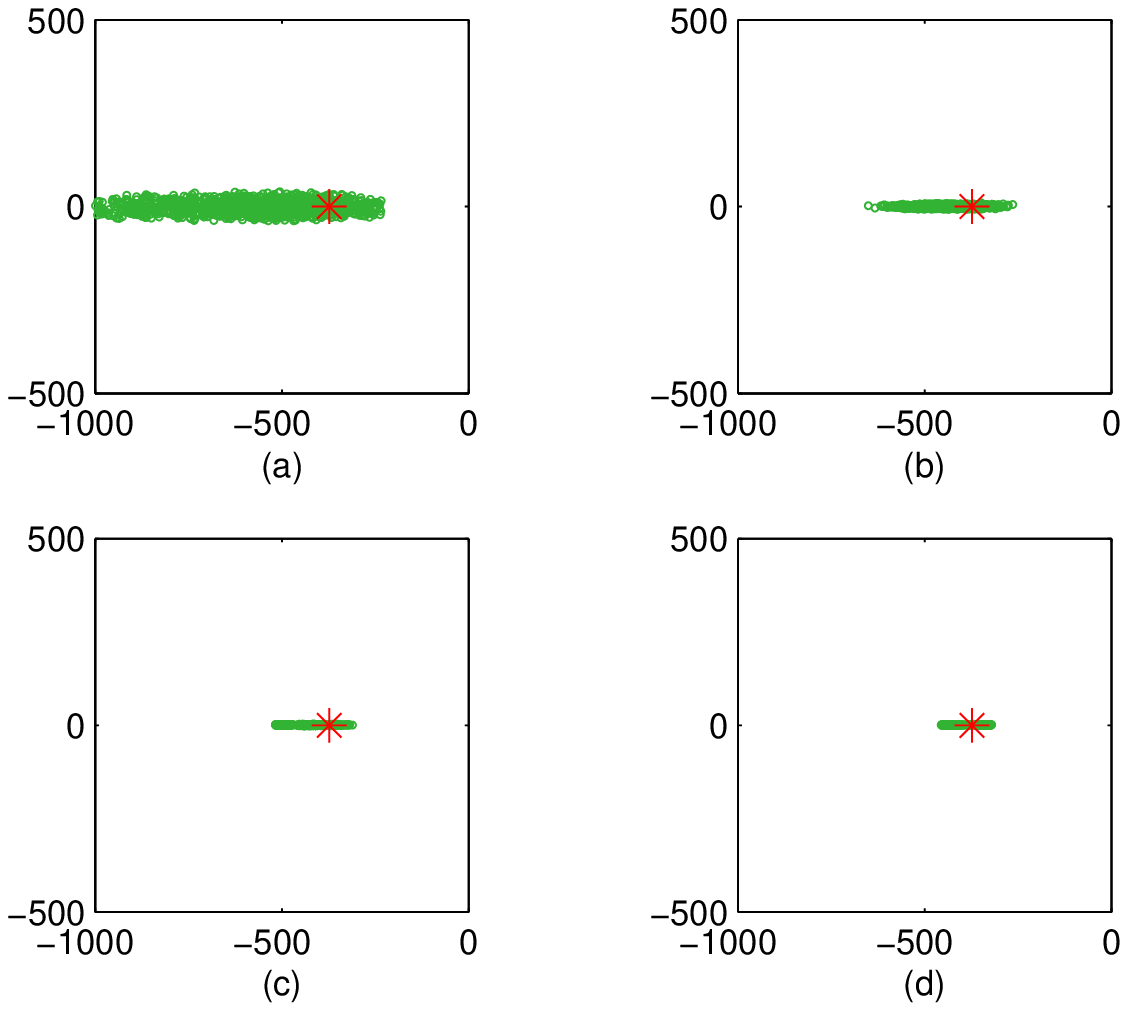}}
 \caption{Scatter plots of random samples $\mathcal{L}^t_m$ in the $(x,y)$ plane (dataset 2).
 The samples approximate the posterior density
$p(\lb|\zb)$. Scatter plots shown after iteration: (a) $t=3$, (b)
$t=6$, (c) $t=9$, (d) $t=12$. The true source location is marked
with a red asterisk. }
 \label{f:6}
\end{figure}

\begin{figure}[htbp]
\centerline{\includegraphics[height=4.5cm]{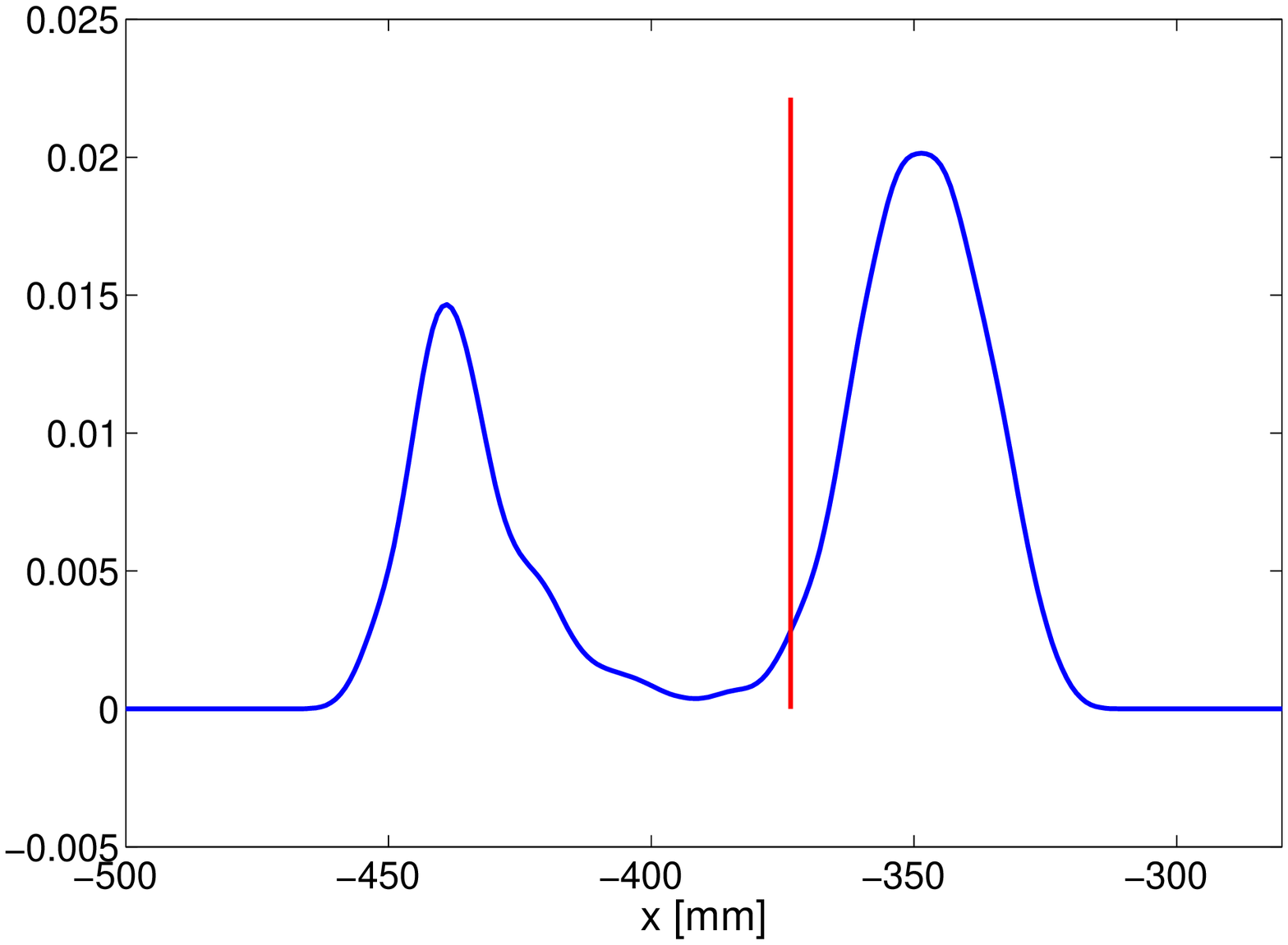}
\includegraphics[height=4.5cm]{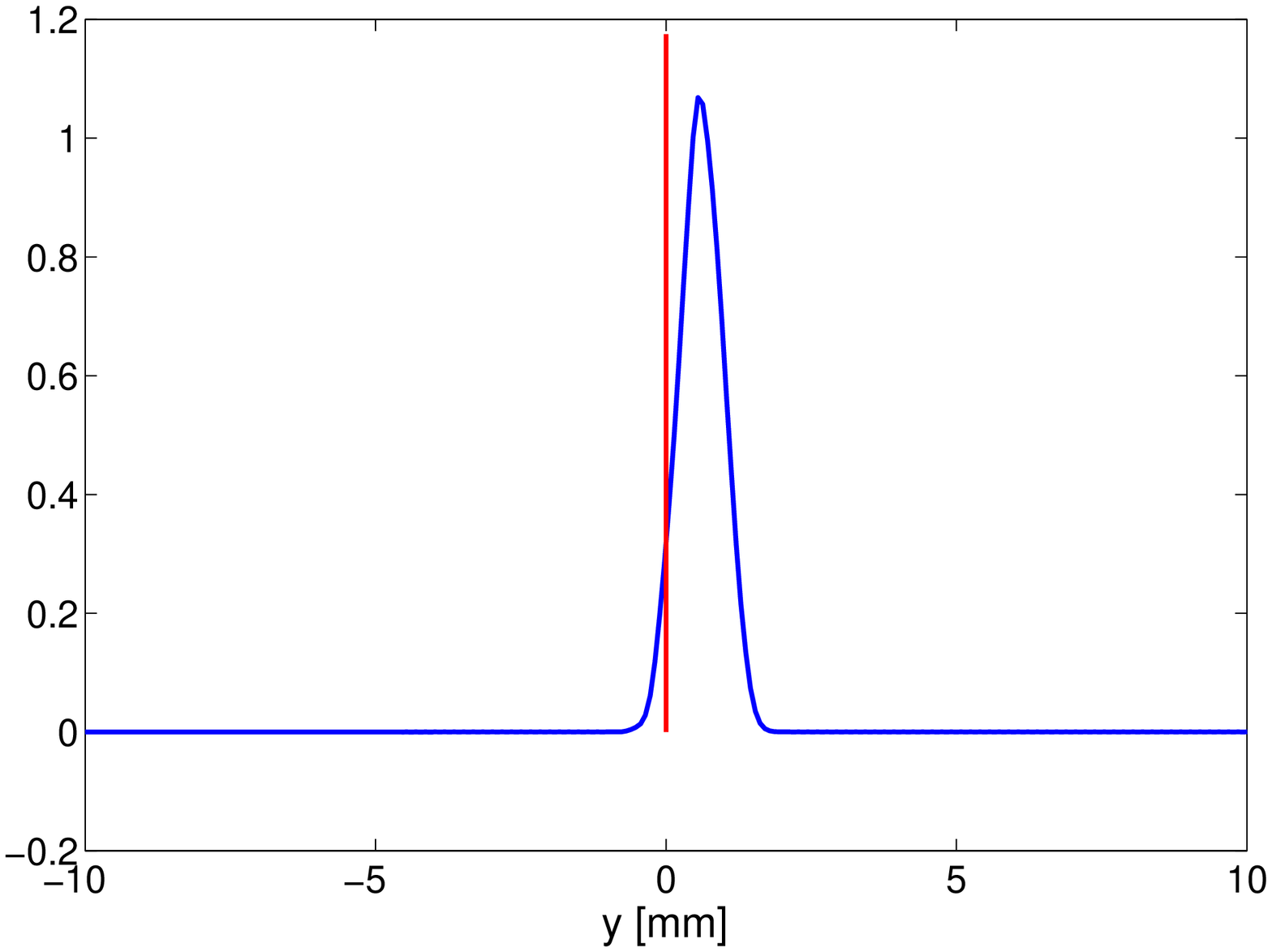}} \centerline{(a)\hspace{6cm}(b)}
 \caption{Estimated posterior density of source location (dataset 2): (a) $x$ axis, (b) $y$ axis.
 The true values are indicated by the vertical solid lines. }
 \label{f:7}
\end{figure}

The dataset 2 was collected in a more challenging setup with
obstacles. The results for dataset 2 are shown in Figs. \ref{f:5},
\ref{f:6} and \ref{f:7}. Fig.\ref{f:5}.(a) shows the computed
tolerance levels over all iterations: note that it took $t=12$
iterations to reach the final tolerance level of
$\epsilon_{\text{\tiny T}} = 2.75\cdot 10^{-9}$.  The average
acceptance rate, for all iterations, was $2.9\%$. The model
probabilities, $p(m|\zb)$ are plotted in Fig.\ref{f:5}.(b). We can
observe that after iterations 7,8 and 9, $m=1$ appears to be the
preferred model. However, at iteration 10,  its probability drops to
zero. As in the case of dataset 1, the probabilities of models $m=2$
and $m=3$ remain non-zero at the lowest tolerance levels. It has
already been noted that stretched exponential solutions may not be
the most appropriate model for these datasets, even though the model
has some physical motivation \cite{yee06}. The analysis presented
here reinforces these findings, and highlights the fact that more
complex, higher-dimensional models are not necessarily better in
describing inherently stochastic phenomena.

Fig.\ref{f:6} displays the scatter plot of random samples
$\mathcal{L}^t_m$  at iterations: (a) $t=3$, (b) $t=6$, (c) $t=9$
and (d) $t=12$. The true source location is marked by a red
asterisk. Once again, localisation of the source  improves with
iterations. However, a close inspection of the final localisation
posterior $p(\lb|\zb)$, shown in Fig.\ref{f:6}.(d), reveals that
this posterior is bi-modal. This can be seen from Figs.\ref{f:7}
which displays the final posterior distribution marginalised to $x$,
and $y$ axes, respectively. The two modes,  clearly seen in
Fig.\ref{f:7}.(a), correspond to models $m=2$ and $m=3$, after
iteration $t=12$. They both appear to be biased: the stronger mode
 is due to model $m=3$ (the probability of this model is higher at $t=12$, see see Fig.\ref{f:5}.(b));
 it peaks  approximately at $\hat{x}_0\approx -350$.

We can make the following observations with respect to the results
obtained using dataset 2. First, none of the models seem to be
correct since the estimation of the $x_0$ coordinate appears to be
biased. Second, the use of multiple-models was beneficial, because
the support of the posterior contains the true source location
(albeit in the tail of one of the modes).

Finally, a remark on the proposed iterative multiple-model ABC
sampler: this algorithm falls into the category of {\em anytime
algorithms} \cite{Zaimag96}, because it returns a valid approximate
solution (i.e. approximate posterior) even if it is interrupted
before the termination criterion is reached.

\section{Summary}
\label{s:6}

The paper presents a robust method for localisation of a biochemical
source. In the absence of an accurate model of the measurement
likelihood, a likelihood-free approximate Bayesian computation (ABC)
approach  is adopted. Furthermore, since there is no universal
dispersion model applicable for all situations (terrain,
meteorological conditions), a multiple-model approach is proposed,
whereby all candidate models are active in parallel and assigned the
probability of being correct. The proposed method computes
adaptively the tolerance levels which are required for ABC sampling.
The method has been tested using two experimental datasets (one in
the open terrain, the other with obstacles) using three dispersion
models (two Gaussian plume models and the stretch exponential
model). Source localisation was very accurate using the open terrain
dataset. The dataset with obstacles presented a significant
challenge, resulting in a bi-modal posterior distribution of source
location. Future work will investigate other dispersion models and
the performance of source localisation using binary sensors.

\bibliographystyle{elsarticle-num}

\small
\bibliography{refs}

\end{document}